\documentclass[sn-vancouver,Numbered]{sn-jnl}
\usepackage{graphicx} % Required for inserting images
\usepackage{lipsum} % for text-filler
\usepackage[title]{appendix}%
\usepackage[dvipsnames]{xcolor} % For uisng colored texts with simple in-line code
\usepackage{textcomp}%
\usepackage{manyfoot}%
\usepackage{multirow}%
\usepackage{booktabs}%
\usepackage{amsmath,amssymb,amsfonts}%
\usepackage{amsthm}%
\usepackage{listings}%
\usepackage{fix-cm}
\usepackage{float}
\usepackage{subcaption}
\usepackage{upgreek}
\usepackage{lineno}
\usepackage{everypage}
\begin{document}
% \linenumbers % Enable line numbering

\AddEverypageHook{\FootnoteA{This document is the unedited Authors’ version of a submitted manuscript currently under review.}}

% \AddEverypageHook{\footnote{Test}}

\title[Article Title]{Imaging current flow and injection in scalable graphene devices through NV-magnetometry}

\author*[1]{\fnm{Kaj} \sur{Dockx}}\email{k.dockx@appliednanolayers.com}
\author*[1]{\fnm{Michele} \sur{Buscema}}\email{m.buscema@appliednanolayers.com}
\author*[2]{\fnm{Saravana} \sur{Kumar}}\email{saravana.kumar@tno.nl}
\author[2]{\fnm{Tijmen} \sur{van Ree}}\email{tijmen.vanree@tno.nl}
\author*[2]{\fnm{Abbas} \sur{Mohtashami}}\email{abbas.mohtashami@tno.nl}
\author[2]{\fnm{Leon} \sur{van Dooren}}\email{leon.vandooren@tno.nl}
\author[2]{\fnm{Gabriele} \sur{Bulgarini}}\email{gabriele.bulgarini@tno.nl}
\author[1]{\fnm{Richard} \sur{van Rijn}}\email{r.van.rijn@appliednanolayers.com}
\author[2]{\fnm{Clara I.} \sur{Osorio}}\email{clara.osoriotamayo@tno.nl}
\author[3]{\fnm{Toeno} \sur{van der Sar}}\email{T.vanderSar@tudelft.nl}


\affil[1]{\orgname{Applied Nanolayers}, \orgaddress{\street{Zilverstraat 1}, \city{Zoetermeer}, \postcode{2718RP}, \country{The Netherlands}}}

\affil[2]{\orgname{Netherlands Organisation for Applied Scientific Research (TNO)}, \orgaddress{\street{Stieltjesweg 1}, \city{Delft}, \postcode{2628CK}, \country{The Netherlands}}}

\affil[3]{ \orgdiv{Department of Quantum Nanoscience}, \orgname{Kavli Institute of Nanoscience},  \orgaddress{\street{Delft University of Technology}, \city{Delft}, \postcode{2628CJ}, \country{The Netherlands}}}

%%===========================================================

% old abstract text: Although graphene and other two-dimensional (2D) materials have been extensively studied for their unique electronic properties and potential for novel electronic devices, several fundamental questions remain about current flow and injection into these materials. We 

\abstract{

The global electronic properties of solid-state devices are strongly affected by the microscopic spatial paths of charge carriers. Visualising these paths in novel devices produced by scalable processes would provide a quality assessment method that can propel the device performance metrics towards commercial use. Here, we use high-resolution nitrogen-vacancy (NV) magnetometry to visualise the charge flow in gold-contacted, single-layer graphene devices produced by scalable methods. Modulating the majority carrier type via field effect reveals a strong asymmetry between the spatial current distributions in the electron and hole regimes that we attribute to an inhomogeneous microscopic potential landscape, inaccessible to conventional measurement techniques. In addition, we observe large, unexpected, differences in charge flow through nominally identical gold-graphene contacts. Moreover, we find that the current transfer into the graphene occurs several microns before the metal contact edge. Our findings establish high-resolution NV-magnetometry as a key tool for characterizing scalable 2D material based devices, uncovering quality deficits of the material, substrate, and electrical contacts that are invisible to conventional methods.
}


\maketitle

%%============================================================

\section{Introduction} \label{sec1}
The past decade has seen a strong effort in the development of graphene and related 2D materials, due to their electronic properties and expected potential for novel devices \cite{review_2D_integrated_circuits, review_advances_graphene_and_TDMCs, review_graphene_and_beyond}. For 2D materials-based electronic devices, the current distribution in the channel and its injection at the contact-channel interface are key to the overall device performance. These quantities are usually studied by macroscopic means, where only spatially-averaged values of relevant Figures of Merit - such as contact resistance - can be extracted. This spatial averaging significantly limits the understanding of the electrical characteristics of 2D materials, where morphological features such as grain boundaries and lattice imperfections play a major role \cite{Tsen2012_grainboundaries, Chen2008_impurity_scattering}. Very recently, nitrogen-vacancy (NV) magnetometry has been used to visualize current flow in graphene-based devices \cite{tetienne2017quantum, Palm2022_bilayerNV, Jenkins2022_ohmNV, Zhong2024_pressureNV, Casola2018_reviewNV, Lee_APL2021, whirl_Palm2024, gFET_Lillie2019}. Most of these studies focus on ground-breaking phenomena such as electron-electron interactions in purposefully-built research devices (e.g. hBN/graphene heterostructures). These devices, while allowing for fundamental understanding of current flow in 2D systems, have little scaling potential and therefore limited relevance from an industrial standpoint.

Here, we use high-resolution scanning NV-magnetometry on scalable graphene-based devices to study the spatial current distribution both in the channel and the contact regions. We find that spatially separated, preferential paths exist for electrons and holes within the same device channel, modulated by the field effect. Moreover, we reveal a strong asymmetry in the electrical behaviour of nominally identical contacts and that current transfers into the graphene several microns before the metal contact edge.
Our findings indicate that certain assumptions that are at the basis of common methods for extracting contact resistance and mobility are not always valid in graphene devices and show that scanning NV-magnetometry is a viable metrology technique for the advancement of novel electronic devices based on 2D materials.

\section{Experimental setup} \label{sec2}
\begin{figure}[!h]
    \centering
    \includegraphics[width=\textwidth]{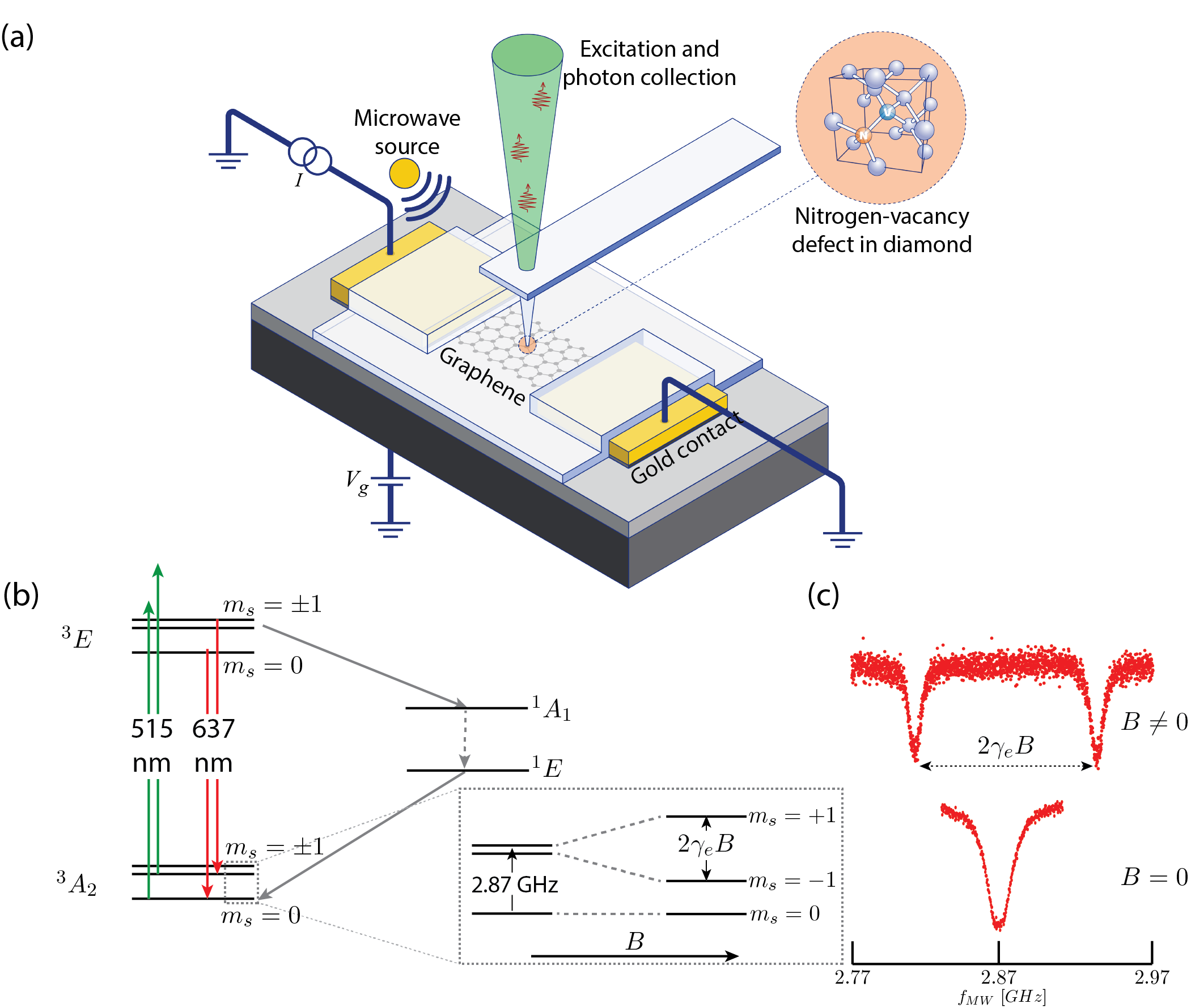}
    \caption{\textbf{Nitrogen-Vacancy (NV) magnetometry: measurement setup and working principles.} (a) A schematic of the experimental setup highlighting the AFM diamond tip containing an NV defect that is excited using a green laser, the microwave antenna providing microwave frequency, and the graphene device with the electrical circuit used to supply voltage for both source-drain bias and gate contacts during the measurements. (b) Energy level diagram of an NV center including the ground-state triplet, the excited-state triplet, and the intermediate singlet states. The ground-state spin levels $m_s=0$ and $m_s=\pm1$ are separated by zero-field splitting ($\approx2.87$ MHz). In the presence of an external magnetic field $B$, the $m_s=\pm1$ energy levels split due to the Zeeman effect. (c) An example of the measured ODMR data (fluorescence as a function of the microwave frequency) in presence of an external magnetic field ($B\neq0$, top curve), and in absence of an external magnetic field ($B=0$, bottom curve). In the absence of external magnetic field, the ODMR curve shows a single dip at a microwave frequency corresponding to the zero-field splitting. In the presence of the magnetic field, the ODMR dip splits in two as a result of the Zeeman effect. The amount of splitting directly corresponds to the strength of the magnetic field which provides the basis for the NV-magnetometry. }
    \label{fig:setup}
\end{figure}

Figure \ref{fig:setup}a shows a schematic of the measurement setup. We first focus our attention to the working principles of the technique after which follows a detailed explanation of the setup. Our quantum-based scanning NV magnetometry setup works on the principle of electron spin resonance. A negatively charged NV center in diamond has a ground-state triplet ($\left|g\right\rangle$, with $^{3}A_{2}$ symmetry and $m_s=0,\ \pm1$), an excited-state triplet ($\left|e\right\rangle$, with $^{3}E$ symmetry and $m_s=0,\ \pm1$), and intermediate singlet state ($\left|s\right\rangle$, with two levels with $^{1}A_{1}$ and $^{1}E$ symmetries) (see Figure \ref{fig:setup}b).  At optical excitation wavelengths below $\sim$640 nm, all spin-states of NV center are photoluminescent for the transition $\left|e\right\rangle\longrightarrow\left|g\right\rangle$. Additionally, for the $m_s=0$ state there also exists a non-radiative path for the $\left|e\right\rangle\longrightarrow\left|s\right\rangle\longrightarrow\left|g\right\rangle$ transition. 

The luminescence rate of the above transitions is affected by external microwave radiation and magnetic fields. Specifically, when a microwave frequency ($\approx2.87$ MHz) resonant with the transition between the spin states is introduced, a reduction in the photoluminescence rate (i.e., a dip in the photoluminescence rate for microwave frequency sweep close to resonance, see Figure \ref{fig:setup}c bottom) is observed due to the change in population within the sub levels of ground and excited states. In the presence of a magnetic field along the NV-axis, the degeneracy between the spin states $m_s=\ +1$ and $m_s=\ -1$ is lifted and the Zeeman splitting takes place (see Figure \ref{fig:setup}b and c top) leading to two dips in the photoluminescence rate. The combination of the above effects is known as the optically detected magnetic resonance (ODMR) signal. The amount of splitting (and thus the resonant frequency shift) is directly proportional to the strength of the magnetic field along the NV-axis. The local magnetic field projected on to the NV-axis can be calculated using $B_{NV}=\left(f_0-f\right)/\gamma_e$, where $f_0$ is the resonant frequency corresponding to a pre-applied bias magnetic field $B_0$ (see below), $f$ is the resonant frequency corresponding to the local magnetic field, and $\gamma_e = 28\,\mathrm{GHz/T}$ is the electron gyromagnetic ratio.

Our experimental setup is built in-house and comprises an atomic force microscopy (AFM) part and magnetometry parts (see Figure \ref{fig:setup}a). The AFM part consists of a modified commercial system (Park XE7), which is used to scan the NV probe over the sample in amplitude-modulated mode. The magnetometry part consists of a \{100\}-cut diamond probe (Qzbare Ltd) with a single nitrogen-vacancy defect that is about 10 nm below the surface of the tip. The sensitivity of the probe in combination of our setup is about $1-10\ \mathrm{\upmu T/\sqrt{Hz}}$. The NV axis is oriented downward and parallel to the direction of the probe with a nominal angle of ${54.7}^\circ$ with respect to the vertical symmetry axis of the tip.  The diamond probe is mounted on one arm of a tuning fork, which is connected to the electronics necessary to enable force feedback during AFM operations. A green ($\lambda$=515nm) laser (Cobolt 06-01 series, output power 120 mW) is used for continuous optical excitation of the NV center. The laser power delivered at the NV center is attenuated using a series of optical filters and controlled via a voltage-controlled attenuator (Thorlabs V450A). A microwave antenna, comprising a 200 micrometer thin loop of copper wire, is positioned next to the diamond probe. The antenna is connected to a microwave signal generator (Rohde \& Schwarz SGS100a) via an amplifier (Mini-Circuits ZHL-5W-422+) to deliver a continuous microwave signal to the probe. An I/Q modulator (Keysight 33622A waveform generator) that is connected to the microwave generator is used to control the sweep of the microwave signal across a desired frequency range. The collected fluorescence of the NV center is measured via an avalanche photodiode module (Excelitas SPCM-AQRH-14-FC). To provide the bias magnetic field ($B_0$) for the NV center, a permanent magnet is placed next to the sample such that its magnetic field is oriented mostly parallel to the NV-axis.

The graphene sample is mounted on the setup as shown in Figure \ref{fig:setup}a. A constant current and a gate voltage are applied to the desired graphene device channel. These signals originate, respectively, from a current source (Aim TTi SMU4201) and a voltage source (Delta Elektronika E0300). The AFM scan rate is kept low (0.01 Hz or 0.02 Hz) to increase the signal-to-noise ratio of the ODMR signals. The laser power at the NV center and the microwave power are set to optimize the sensitivity of each of the NV probes that are used in the experiments\cite{dreau2011avoiding}. The mapping of the frequency variations of the ODMR data, and consequently the magnetic field variations of the sample, are obtained with the following procedure. While the NV probe is continuously scanned over the sample, the microwave frequency is continuously and independently swept over the frequency range of the ODMR dip, resulting in a set of position dependent ODMR data. The microwave frequency sweep rate is set such that there are at least two sweeps per scanned AFM pixel. The AFM scans and the microwave sweeps are synchronized using a time tagger (Time Tagger 20, Swabian Instruments) which is also used to record the photon count data as a function of time. These time series signals are then subjected to an adjusted frequency-domain low-pass filtering to reduce the noise and are subsequently fitted with a second-order polynomial to find the microwave resonant frequency for each sweep. The extracted resonance frequencies are then averaged accordingly and converted into a single value of microwave resonance frequency ($f$) per AFM pixel.  Each AFM scan (and thus the magnetic and current density map) is $50~\upmu \mathrm{m} \times 50~\upmu \mathrm{m}$ in size with $100 \times 100$ pixels (unless otherwise specified). Larger scans are obtained by using cross-correlation of the AFM image data to stitch individual scans together. The result is a map of highly localized magnetic field strength.

Ultimately, a current density map can be reconstructed from the 2D magnetic field strength map using Bio-Savart’s law and the continuity equation in Fourier space \cite{broadway2020improved,tetienne2017quantum}. 
Given that reconstructing current from the collected data is non-trivial, we have calibrated the procedure on a gold strip with known transport properties. The calibration results are presented in Supplementary Figure \ref{supfig: Gold}. Moreover, our conclusions are independent of the absolute current density values. Instead, we focus on its spatial distribution. Therefore, small and linear inaccuracies in the current extraction procedure do not affect our findings. 

\newpage

\subsection{Graphene sample characteristics} \label{subsec2-1}
\begin{figure}[H]
    \centering
    \includegraphics[width=0.75\textwidth]{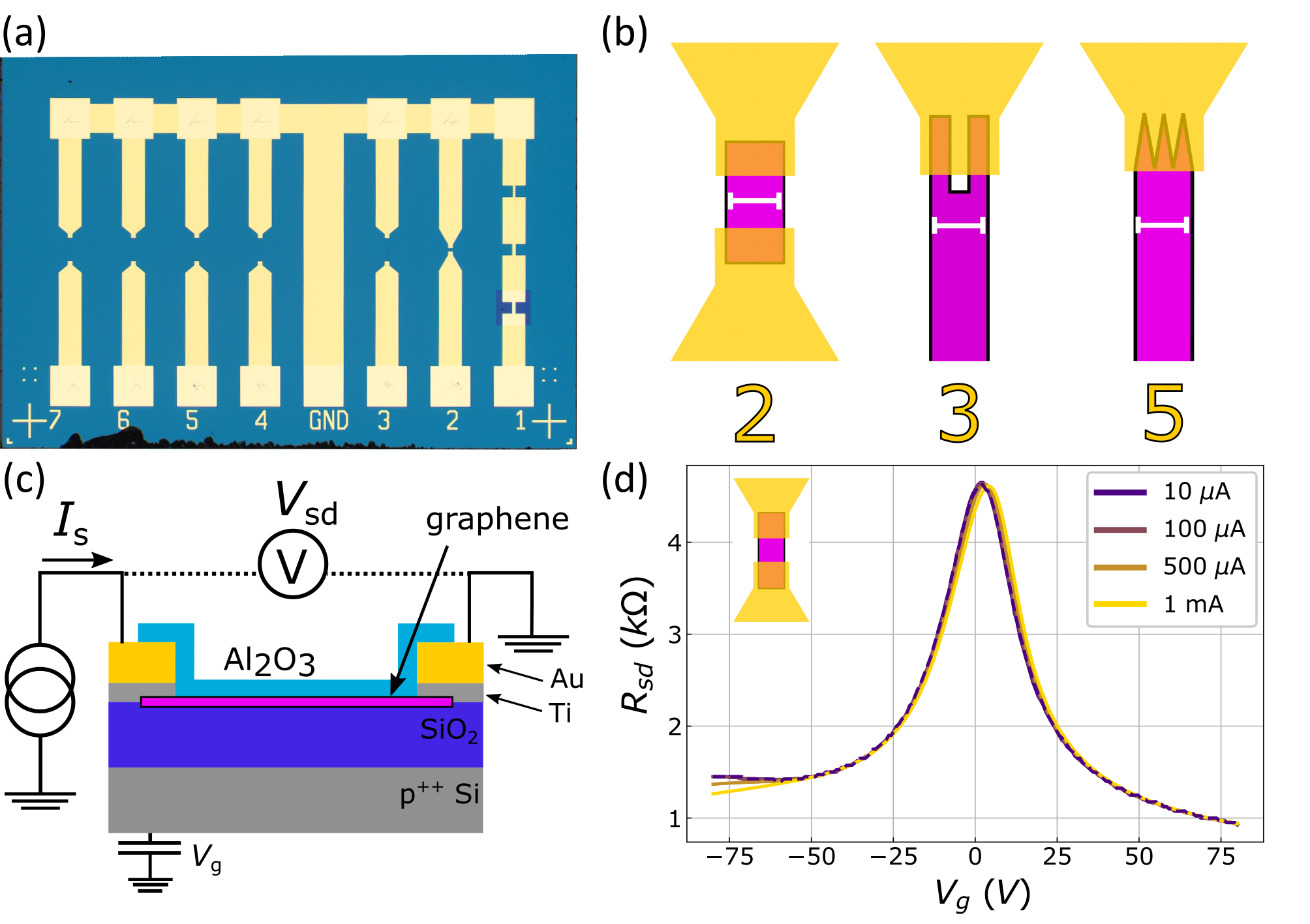}
    \caption{\textbf{Graphene sample: design, fabrication, and electrical characterisation.} (a) Micrograph of a sample highlighting the common drain (GND) and the seven channels of which one features a gold test structure (channel 1) and six feature graphene field-effect transistors (gFETs; at channels 2 to 7). (b) Top-view schematics of (part of) the gFETs on channels 2, 3 and 5. Scale bars are 30 $\upmu$m. (c) Cross-section schematic of a gFET and circuit used for electrical characterisation. Monolayer graphene is transferred to a $\mathrm{Si/SiO_2}$ substrate, the channel is defined and contacts are patterned on top. The devices are encapsulated with 20 nm of $\mathrm{Al_2O_3}$ using atomic layer deposition. To obtain a transfer characteristic, a constant source current, $I_s$, is applied to the device while the source-drain voltage, $V_{sd}$, is monitored as a function of gate voltage, $V_g$. (d) Two-terminal transfer characteristics for increasing source currents of gFET 2. Dotted indigo line represents a second measurement at 10 $\upmu$A bias taken after the high current measurements.}
    \label{fig:sample}
\end{figure}

The samples subject to our study contain graphene field-effect transistors (gFETs) fabricated on a 4-inch wafer using scalable techniques: chemical vapor deposition (CVD) for graphene growth, dry transfer process, and photolithography. We leverage scalable fabrication methods to validate NV-magnetometry as a meaningful metrology tool for advancing 2D devices. 

Figure \ref{fig:sample}a shows one of the measured samples, composed of 7 separate channels and one common drain (GND). Channels 7 through 2 contain gFETs of various geometries, while channel 1 features gold structures for testing and calibrating the measurement setup. The placement of the gold pads enables automated measurements in a probe station, facilitates the wire-bonding of the graphene sample to a PCB, and also allows the NV tip to approach the sample without geometric interference from the bond wires.   

In the following, we focus our attention on gFETs 2, 3 and 5 (located at channels 2, 3, and 5), whose shape and size are sketched in Figure \ref{fig:sample}b. The graphene geometry at the metal-graphene interface differs significantly between these devices. The geometries are purposely designed to study how the current is injected from the metal contacts into the graphene channel. GFET 2, with a size of $30~\upmu \mathrm{m} \times 30~\upmu \mathrm{m}$ and a contact region (defined as the overlapping area between the gold and graphene) measuring $30~\upmu \mathrm{m} \times 30~\upmu \mathrm{m}$ on each side, allows us to measure the current flow through the entire device within reasonable time constraints. Due to the fact that gFETs 3 and 5 are longer, we focus our attention at their contact region. GFET 3 features two parallel graphene strips of equal dimensions at the metal/graphene interface: we refer to this device as ``two-finger prong''. Using this geometry, we can directly compare the current injection between two graphene contacts within a single image. This is to assess whether these graphene patches, nominally identical in fabrication flow and size, also exhibit identical electrical behaviour. The contact area of gFET 5 is designed with a ``saw-tooth" pattern to investigate where the current injection occurs along the graphene contact length.  

A cross-section schematic of the gFETs is depicted in Figure \ref{fig:sample}c, along with the measurement circuit used to characterize the gFETs electrically. Briefly, the device fabrication is as follows: a monolayer CVD graphene is transferred to a heavily p-doped 4-inch silicon wafer with a dry, thermal 285 nm $\mathrm{SiO_2}$ layer. The Si is used as back gate while the $\mathrm{SiO_2}$ serves as the gate dielectric. The graphene layer is then patterned via reactive ion etching (RIE), and metal contacts (5 nm Ti + 50 nm Au) are deposited on top. The device channels and contacts are defined by standard photolithography processes. The devices are encapsulated with a 20 nm Al$_2$O$_3$ layer applied by atomic layer deposition (ALD). The ALD layer serves multiple functions: (1) it protects the graphene during magnetometry measurements by avoiding direct contact between the NV-tip and the graphene surface; (2) it prevents environmental degradation of the graphene; and, more importantly, (3) it reduces the doping of the graphene devices, resulting in a Dirac point position near zero gate voltage \cite{Dockx2024}. For more details on the fabrication procedure we refer to reference \cite{Dockx2024}. 

Before executing the magnetometry measurements, the graphene devices are characterised electrically to extract key properties, such as the charge neutrality point ($V_{cnp}$), i.e. the gate voltage ($V_g$) value at which the source-drain resistance ($R_{sd}$) reaches its maximum. The device transfer curves are obtained by sweeping $V_g$ on the Si substrate while applying a constant source current ($I_s$) and measuring the source-drain voltage ($V_{sd}$). Figure \ref{fig:sample}d shows the measured transfer curves for gFET 2, plotted as $R_{sd} = \frac{V_{sd}}{ I_{s}}$ versus $V_g$ for four different bias currents ranging from 10 $\upmu$A to 1 mA. The curves show that the $V_{cnp}$ of the device is located close to 0 V (as a result of the ALD encapsulation \cite{Dockx2024}) and remains unchanged for different $I_s$ values. This allows us to easily access both the hole ($V_g < V_{cnp}$) and the electron ($V_g > V_{cnp}$) transport regimes. High currents of $\sim$1 mA are required for the magnetometry measurements to achieve sufficient resolution in our setup. By applying source currents from 10 $\upmu$A to 1 mA, we confirm that the device exhibits stable behaviour at the conditions required for NV-magnetometry. After applying increasingly higher $I_s$ up to 1mA, we remeasured the transfer curve using the typical, low $I_s$ of 10 $\upmu$A (see dotted indigo line in Figure \ref{fig:sample}d). Since no significant difference is observed between these curves, we can conclude that high $I_s$ do not permanently change the device behaviour. 

\section{Results} \label{sec3}

\begin{figure}[H]
    \centering
    \includegraphics{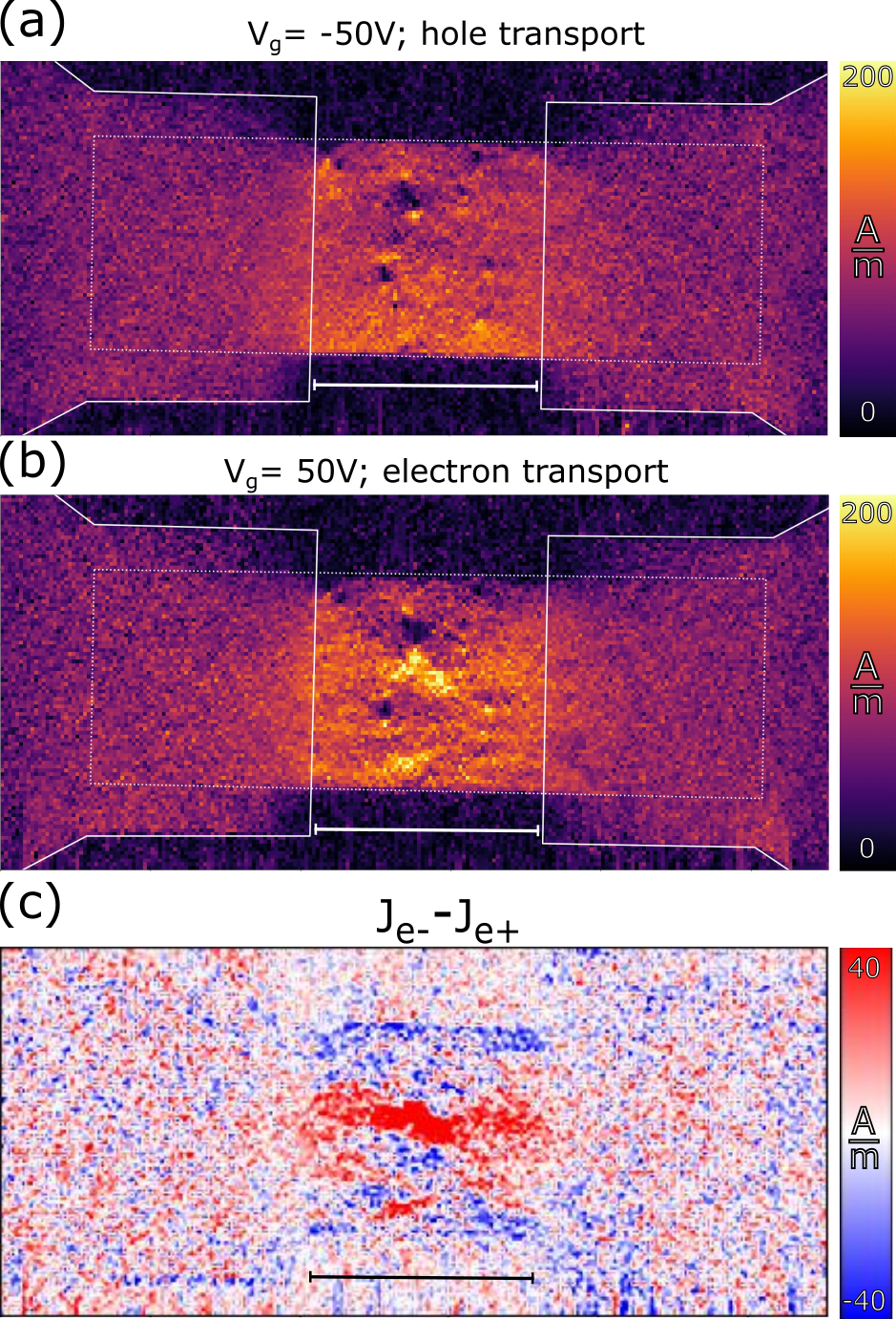}
    \caption{\textbf{Current imaging in the hole and electron transport regime.} (a)-(b) Current density maps of a $30~\upmu \mathrm{m} \times 30~\upmu \mathrm{m}$ graphene channel (gFET 2) at varying gate voltages and transport regimes: (a) $V_{g} = -50 \ $V: hole regime. (b) $V_{g} = +50 \ $V: electron regime. The shown current densities are a sum of two measurements performed at +1.5 mA and $-$1.5 mA of bias current. (c) False color map showing the difference between electron and hole current densities ($J_{e-}$ and $J_{e+}$).
    }
    \label{fig:channel2}
\end{figure}

The results of the magnetometry measurements performed on gFET 2 are presented in Figure \ref{fig:channel2}. Panels (a) and (b) show false color maps of the current density throughout the entire device at different doping regimes. Each image is constructed by stitching together three separate measurements obtained consecutively on adjacent areas of the device. The plotted current density for each section is derived from a single measurement, during which the AFM scans each line of the image twice: once in the forward direction and once in the backward direction. In the forward scan, a current of +1.5 mA is applied, while in the backward scan, the current is set to $-$1.5 mA. For each pixel, the measured magnetic fields corresponding to the forward and backward currents are subtracted to calculate the change in the magnetic field, which is then used to construct the current density maps.

The boundaries of the contacts (solid white line) and the graphene patch (dotted white line) in Figure \ref{fig:channel2}(a) and (b), are extracted from the simultaneously acquired AFM amplitude error data and are overlayed to the magnetometry data to illustrate the corresponding edges. For reference, the amplitude error data of the measured devices are presented in Supplementary Figure \ref{supfig: amplitude_error}.

The applied gate voltage in Figures \ref{fig:channel2}(a)-(b) coincides with different transport regimes: panel (a) is acquired at $V_{g}$ = $-$50 V, at a hole carrier density $n \approx -3.8\cdot10^{12}\mathrm{cm^{-2}}$ while panel (b) is acquired at $V_{g}$ = +50 V, resulting in an electron carrier density $n \approx +3.8\cdot10^{12}\mathrm{cm^{-2}}$. Panels (a) and (b) show current inhomogeneities of two major kinds: common ``dark spots" and independent ``hot spots". The dark spots are regions where the current density is at least ten times lower than the maximum, and likely coincide with physical defects in the graphene channel\cite{tetienne2017quantum, Zhong2024_pressureNV}. All identifiable physical defects in the SEM images (Supplementary Figure \ref{supfig: SEM}) correspond to dark spots in the current density maps.
On the other hand, the hot spots ($J >$ 150 A/m) change position in the device channel as the gate voltage is varied, suggesting the presence of low resistive pathways that depend on the majority carrier type. These pathways can be a result of local charge carrier inhomogeneities (charge puddles) \cite{zhang2009origin} that are linearly superimposed over the global gate effect. In other words, p-type charge puddles will contribute more holes for transport when the gate biases the entire channel into the hole regime, increasing the current density in those spots. The same is true for n-doped puddles in the electron regime. Inhomogeneous distribution of charged residues, such as resists and charged impurities in the oxides ($\mathrm{SiO_2}$ and/or $\mathrm{Al_2O_3}$), can cause spatial variations in doping across the channel \cite{zhang2009origin}. 

In more detail, with electrons as the majority charge carriers (Fig. \ref{fig:channel2}b), the current predominantly flows through the center of the channel. In contrast, the current prefers to flow along the device edges when holes are the majority carriers (Fig. \ref{fig:channel2}a). This is highlighted in Figure \ref{fig:channel2}c, where a differential map is shown, generated by subtracting the hole current density ($J_{e+}$) from the electron current density ($J_{e-}$). In Supplementary Figure \ref{supfig:Hall_effect} we exclude the Hall effect being the cause of the measured spatial separation between the electron and hole current pathways. Other reports (e.g.  S. E. Lillie et. al \cite{gFET_Lillie2019}) speculate that device degradation can influence the observed current pathways. To exclude degradation effects we performed repeated electrical measurement and repeated magnetometry measurement that show minimal change throughout the measurement campaign (see Supplementary Figures \ref{supfig: EL_stability} and \ref{supfig: repeatability}). A likely explanation for the difference in findings is that our fabrication process preserves the graphene integrity. Therefore we can conclude that the observed inhomogeneities in Figure \ref{fig:channel2} arise from an inhomogeneous potential landscape driven by spurious charges in close proximity to the graphene channel (residues, impurities in the oxides, etc.).

Understanding the contact region is of paramount importance for the development of 2D devices \cite{TSMC_TMDC, IMEC_TMDC_IEDM, IEEE_Intel_TMDC, TSMC_contacts_Wu2024}. In the following, we leverage the unparalleled ability of NV-magnetometry to non-invasively image current densities through stacks of different materials to study the current injection at the graphene-gold interface. 

% probe inside contacts, the devices at channels 3 and 5 (see Fig. \ref{fig:channel2}c) have been designed with different contact geometries. These variations can provide further insights into current injection mechanisms from metals into DOS limited channel materials such as graphene. 

\begin{figure}[H]
    \centering
    \includegraphics[width=0.75\textwidth]{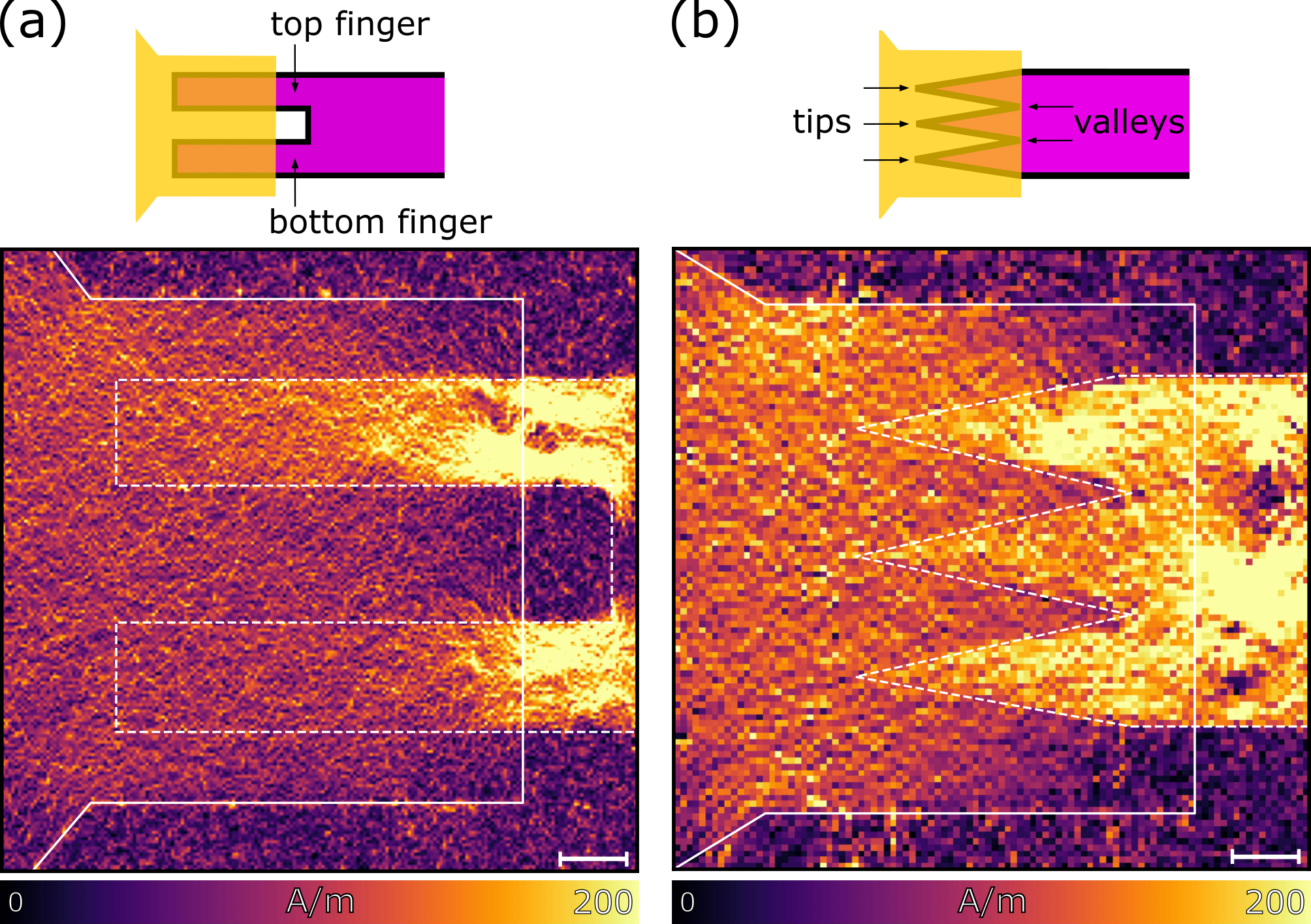}
    \caption{\textbf{Current imaging at the gold-graphene interface.} Current density maps corresponding to (a) two-finger prong graphene structure (gFET 5), and (b) saw-tooth graphene structure (gFET 3). Scale bars are 5 $\upmu$m. Each current density map is a result of two measurements performed at +1 mA and $-$1 mA of bias current. Above each density map a schematic is shown highlighting the contact geometries with their relevant nomenclature.}
    \label{fig:contact_geometry}
\end{figure}

To investigate how contact geometry affects the current injection in our gFETs, we measure the current density in the contact region of gFETs 3 and 5. While gFET 2 has a typical contact geometry (i.e. a graphene rectangle underneath a metal patch), gFET 3 and gFET 5 feature a structured contact geometry: a saw-tooth for gFET 3 and a two-finger prong for gFET 5. Figure \ref{fig:contact_geometry} shows the current density maps of the top contact area of each of the devices: the two-finger prong in Figure \ref{fig:contact_geometry}a (gFET 5) and the saw-tooth in Figure \ref{fig:contact_geometry}b (gFET 3). A top view schematic of each contact area is visible above the density maps along with the corresponding nomenclature.  

The two-finger prong in Figure \ref{fig:contact_geometry}a allows for a direct comparison between two contacts that are identical in design and have been fabricated in the same step. These conditions are usually assumed to give rise to electrically equivalent contacts. In stark contrast to the above expectations, the two fingers do not carry equal currents. The current density in the top finger is significantly higher compared to the bottom one. Even the likely presence of a defect in the middle of the top finger does not seem to diminish the current carried by this side of the contact. Furthermore, we observe that the two fingers appear to have different transfer lengths: the current density in the top finger starts to increase from roughly halfway along its length, while for the bottom finger, the current density begins to increase only just before the contact edge. The above two effects corroborate the idea that the top finger has a lower contact resistance. As more current is injected into the top finger, current injection saturates close to the contact-channel edge and the transfer length increases. The current appears to strike a balance between the more resistive contact at the bottom finger and the added resistance of an increased transfer length in the top finger. The data in Figure \ref{fig:contact_geometry}a strongly suggests that one cannot assume that contacts with the same geometry and fabricated in the same lithographic step have identical electrical behaviour. A possible mechanism for this is a non-uniform thickness of resist residues on the 2D material contact patch which can lead to different injection barriers. This highlights the importance of NV-magnetometry as a metrology technique to progress the development of novel 2D devices, both on a research as well as industrial scale, e.g. by evaluating contact quality and uniformity in 2D devices.

The observation that nominally identical contacts do not have identical current injection capabilities invites a careful evaluation before using common methods to extract contact resistance. In particular, the transfer-length method (TLM) relies on the assumption that each contact pair has an identical contact resistance \cite{TLM_ref}. Our data strongly suggests that this assumption is difficult to satisfy without particular care of the metal-2D material interface. NV-magnetometry can therefore significantly help to optimize a scalable fabrication process yielding uniform device performance.
% This is crucial when developing devices for industrial applications and invites a re-evaluation of methods aimed at extracting contact resistances. 

% The transfer-length method (TLM), for example, is a popular and widely used approach to extract the contact resistances of graphene (and other 2D) devices. However, the TLM method assumes a constant contact resistance and channel resistivity across multiple graphene strips of increasing length. From figure \ref{fig:contact_geometry}a) it is immediately clear that two identically designed contacts do not have identical transport properties. Furthermore, since transport within the graphene channels is influenced not only by physical defects but also by charge puddles, it is implausible that the channel resistivity across a TLM device remains constant. This highlights the limitations of the TLM method in fully characterizing graphene-metal contact resistance. NV magnetometry offers a novel approach to 

For the saw-tooth shaped contact shown in Figure \ref{fig:contact_geometry}b we notice that the graphene teeth carry significantly more current than the surrounding gold. This is, on a first glance, surprising: since gold has a very low resistivity compared to graphene, one would expect the current to stay in the gold as long as possible before transferring into the more resistive graphene layer. In this case, the bulk of the current would be injected near the valleys of the saw-tooth. However, Figure \ref{fig:contact_geometry}b reveals a different picture: the current prefers the graphene layer to the gold from the start. The gold near the valleys of the graphene teeth carries significantly less current compared to the surrounding graphene and compared to the gold near the tips. This indicates that the current is injected along the entire length of contacted graphene. In this case, the transfer length appears to be equal to, or greater than, the entire length of the graphene teeth (= 30 $\upmu$m from the tip of the teeth to the contact-channel edge). Both contact geometries show strong similarities in current injection behaviour, namely inhomogeneity (between the different teeth) and long transfer lengths (also in the fingers). 

Note that the high bias current may cause all regions where current is injected to become saturated. Studying current injection at lower bias current is an interesting topic for future experiments. To enable these measurements, a significant improvement in the sensitivity is required, which is beyond the scope of this paper. Alternatively, with the current setup, one could compare devices with the same shape but different aspect ratio's (e.g. elongated teeth) to extract transfer lengths.

\section{Conclusion}\label{sec5}
We employed high-resolution scanning NV-magnetometry to investigate the spatial distribution of the current in scalable graphene-based devices, focusing on both the device channel and contact regions. Our data reveals the presence of spatially distinct, preferential conduction paths for electrons and holes within the same graphene device, with their behavior modulated by the field effect. Additionally, we observe significant asymmetry in the electrical response of nominally identical contacts and current transfer occurring tens of microns before the geometrical metal edge. We believe our findings are universal, device-independent, and naturally applicable to other 2D devices. As a result, we suggest that conventional approaches to extracting contact resistance may not fully apply to 2D materials. Furthermore, this work highlights scanning NV-magnetometry as a powerful metrology tool offering unparalleled capabilities to advance our knowledge of 2D semiconductor devices. 

\section{Acknowledgments}

This work was supported by the Dutch Research Council (NWO) under the project  NWA.1160.18.208. TNO thanks Sidney Cadot and Joris van Rantwijk from Jigsaw B.V. for developing data acquisition software for the measurement setup. TNO also acknowledges funding from the European Union’s Horizon Europe research and innovation program under grant agreement No 101113901 and from the Dutch Ministry of Economic Affairs and Climate Policy (EZK), as part of the Quantum Delta NL program.

\section{Author contributions}

K.D. and M.B. designed, fabricated and characterized (EL, SEM, AFM) the graphene devices. A.M., K.D., M.B., L.v.D., C.I.O. and T.v.S. designed the experiments, and S.K., T.v.R, L.v.D. and A.M. conducted the experiments. K.D., M.B., S.K., T.v.R and A.M. contributed to the interpretation of the data and wrote the manuscript with comments by all authors. R.v.R., C.I.O., G.B. and T.v.S. initiated the project, secured funding and globally supervised the project.

\newpage
\bibliography{references}

\newpage
\section*{Supplementary Figures}
\subsection*{Gold reference measurement}
\begin{figure}[H]
    \centering
    \includegraphics[width=\linewidth]{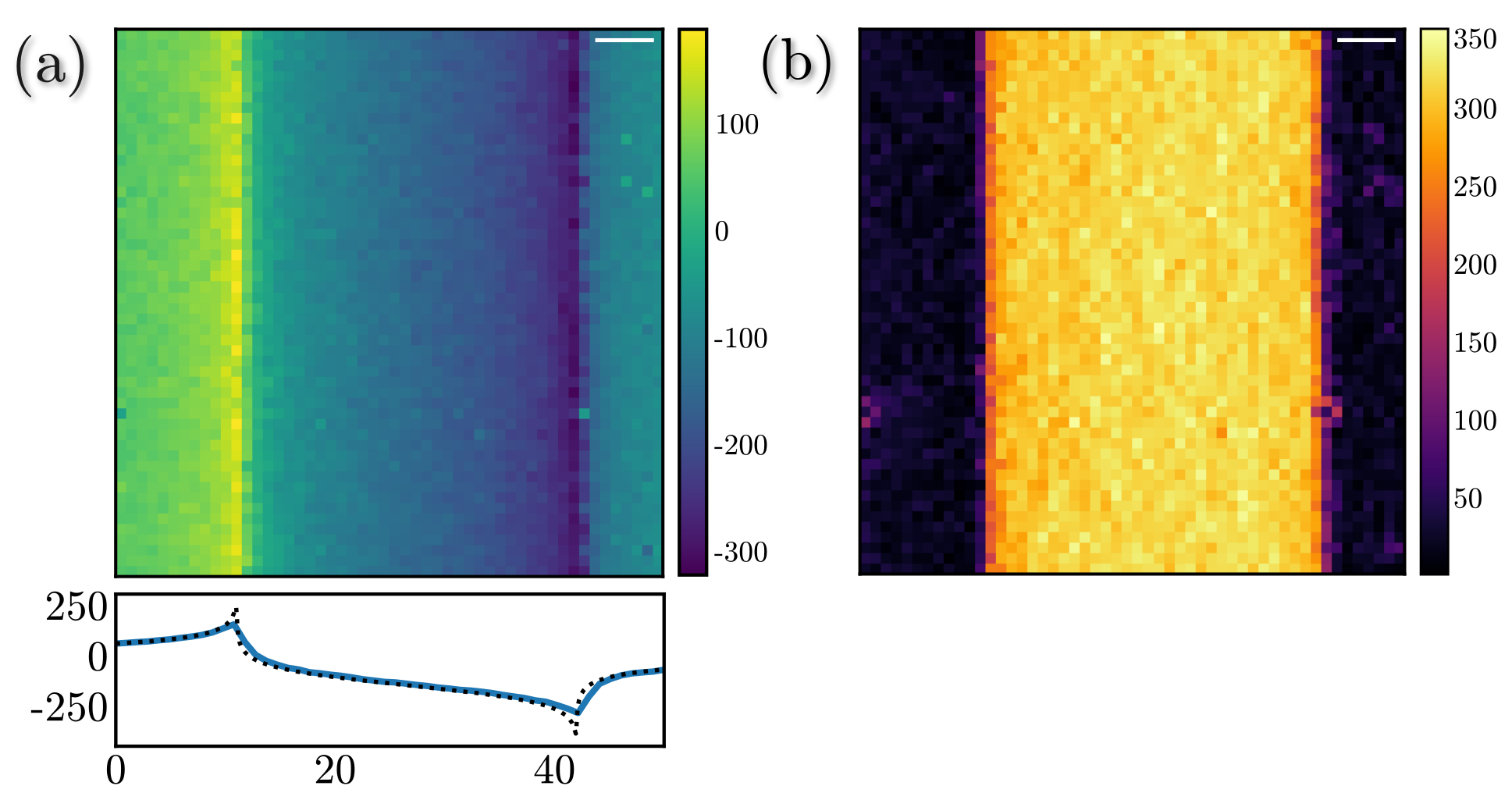}
    \caption{\textbf{Reference NV-magnetometry measurement on gold.} (a) measured magnetic field over gold strip in $\upmu$T (above) for a total current  of 10 mA and its comparison to the theoretical value (below). (b) Corresponding current density map, in units of A/m.}
    \label{supfig: Gold}
\end{figure}

Figure \ref{supfig: Gold} shows a reference magnetometry measurements performed with our setup on the gold strip in channel 1 (Figure \ref{fig:sample}a). A constant current of $\pm5\ $mA is passed through the gold strip and the measured magnetic field is shown in Figure \ref{supfig: Gold}a (top). The measured values are averaged along the axis of the strip as the magnetic field is uniform along this axis and compared with the calculated values (Figure \ref{supfig: Gold}a (bottom)). A very good agreement between the measured and calculated values is observed, which stands as a validation of our measurement procedure. (see also the current density map \ref{supfig: Gold}b)
\newpage
\subsection*{AFM amplitude error}
\begin{figure}[H]
    \centering
    \includegraphics[width=0.8\linewidth]{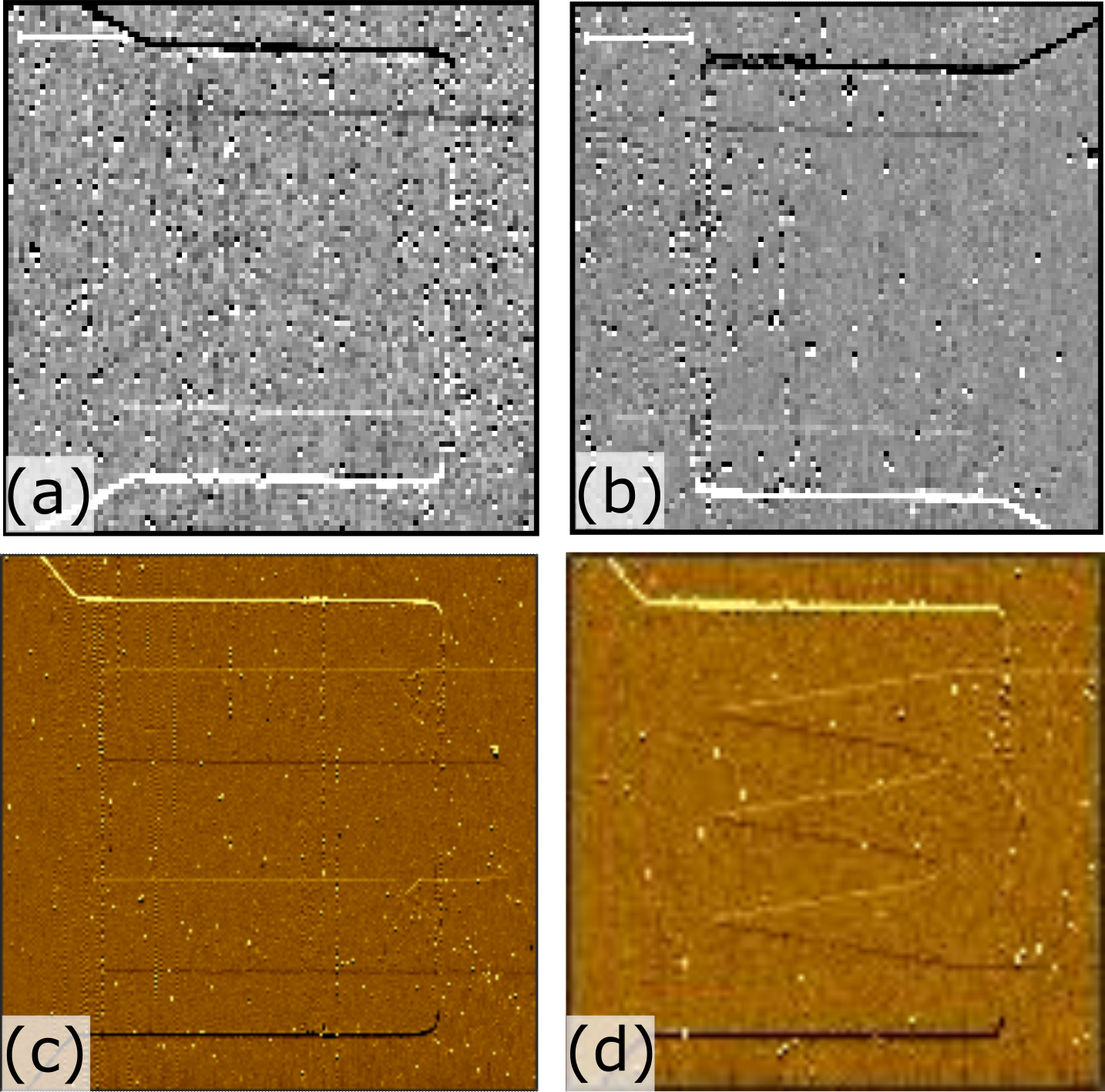}
    \caption{\textbf{AFM amplitude error images acquired during NV-magnetometry measurements.} AFM amplitude error images corresponding to: (a) left contact and (b) right contact of gFET 2, featured in Figure \ref{fig:channel2}. (c) contact area of gFET 5, featured in Figure \ref{fig:contact_geometry}a. (d) contact area of gFET 3, featured in Figure \ref{fig:contact_geometry}b. These data are used to infer contact and channel locations on the current density maps in the main text that are shown by white solid and dotted lines.}
    \label{supfig: amplitude_error}
\end{figure}

Figure \ref{supfig: amplitude_error} shows the amplitude error (difference between set-point amplitude and measured amplitude of the probe) acquired during the NV-magnetometry measurements of the devices in the main text. This data provides information regarding the topography and therefore it is used to identify the contact and channel locations. 
\newpage
\subsection*{Hall effect}
\begin{figure}[H]
    \centering
    \includegraphics[width = 0.9\linewidth]{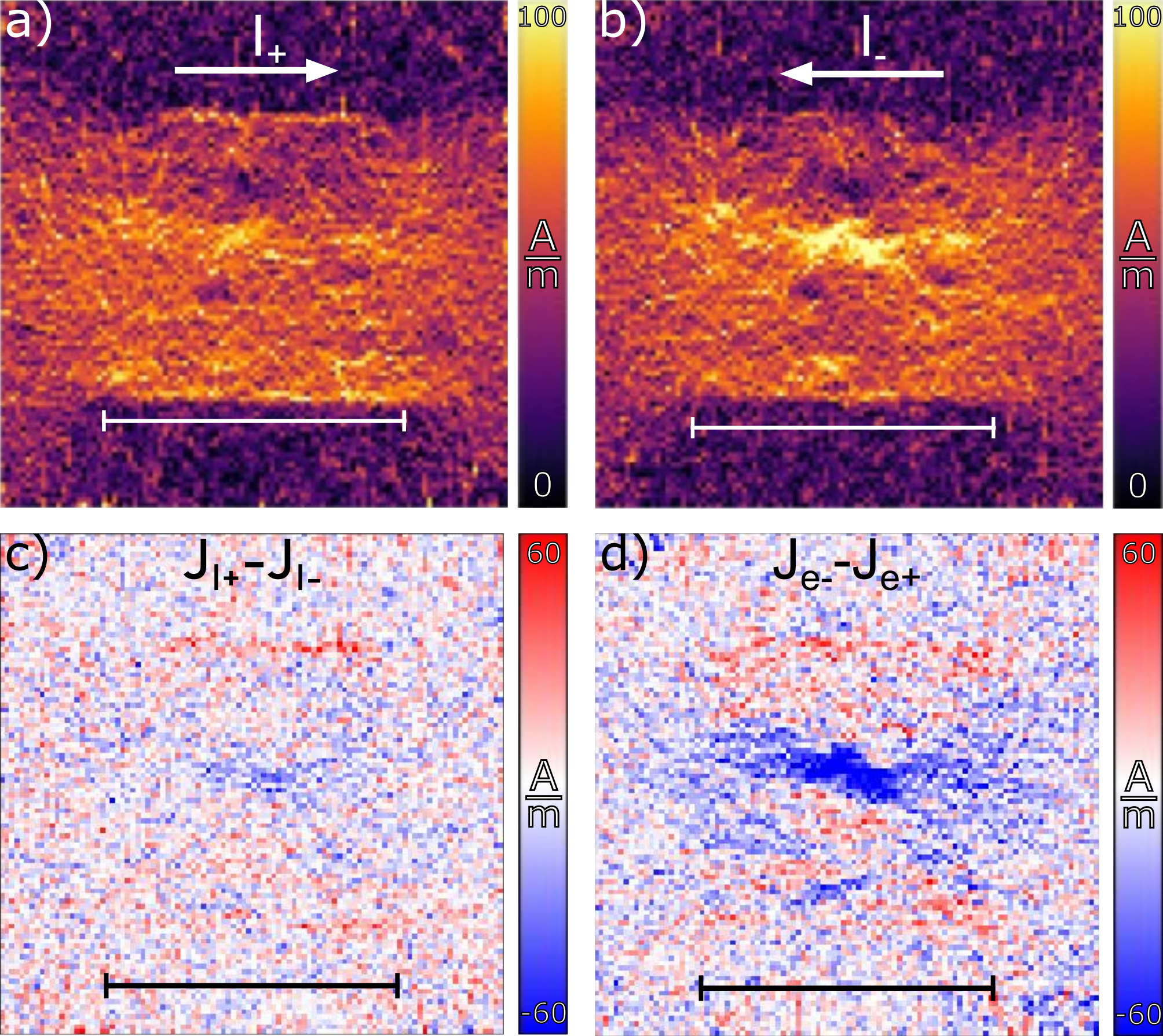}
    \caption{\textbf{Effect of switching current direction on the measurement results.} (Differential) current density maps of gFET 2, featured in Figure \ref{fig:channel2}: (a) current density map for $I = 1\ $mA at $V_g = 0\ $V; (b) current density maps for $I = -1\ $mA and $V_g = 0\ $V. (c) differential density map between positive and negative currents. (d) differential density map between electron and hole transport for scale comparison. All scale bars are 30 $\upmu$m. }
    \label{supfig:Hall_effect}
\end{figure}

In this section we look at the impact of the external constant bias magnetic field needed for the NV-magnetometry, on the current density measurement results. An external magnetic field influences charge carriers through the Lorentz force, causing their deflection depending on the carrier type and current direction. This effect can potentially influence the distribution of the current in the graphene channel. Here, by comparing density maps for opposing current directions under identical conditions, we establish an upper bound on the influence of the magnetic field on our measurements. Figure \ref{supfig:Hall_effect}a and \ref{supfig:Hall_effect}b show density maps of gFET 2  (Figure \ref{fig:channel2}) for opposite current directions ($I_{+}$ and $I_{-}$). All other parameters such as gate voltage and absolute value of the bias current are kept identical. Figure \ref{supfig:Hall_effect}c presents a differential map illustrating the difference between the two current directions. Comparing this map with the differential map for opposite charge carriers in Figure \ref{supfig:Hall_effect}d reveals that any Hall-effect contribution is minimal compared to other factors influencing current flow in the devices, such as charged impurities.
\newpage
\subsection*{SEM}
\begin{figure}[H]
    \centering
    \includegraphics[width=\linewidth]{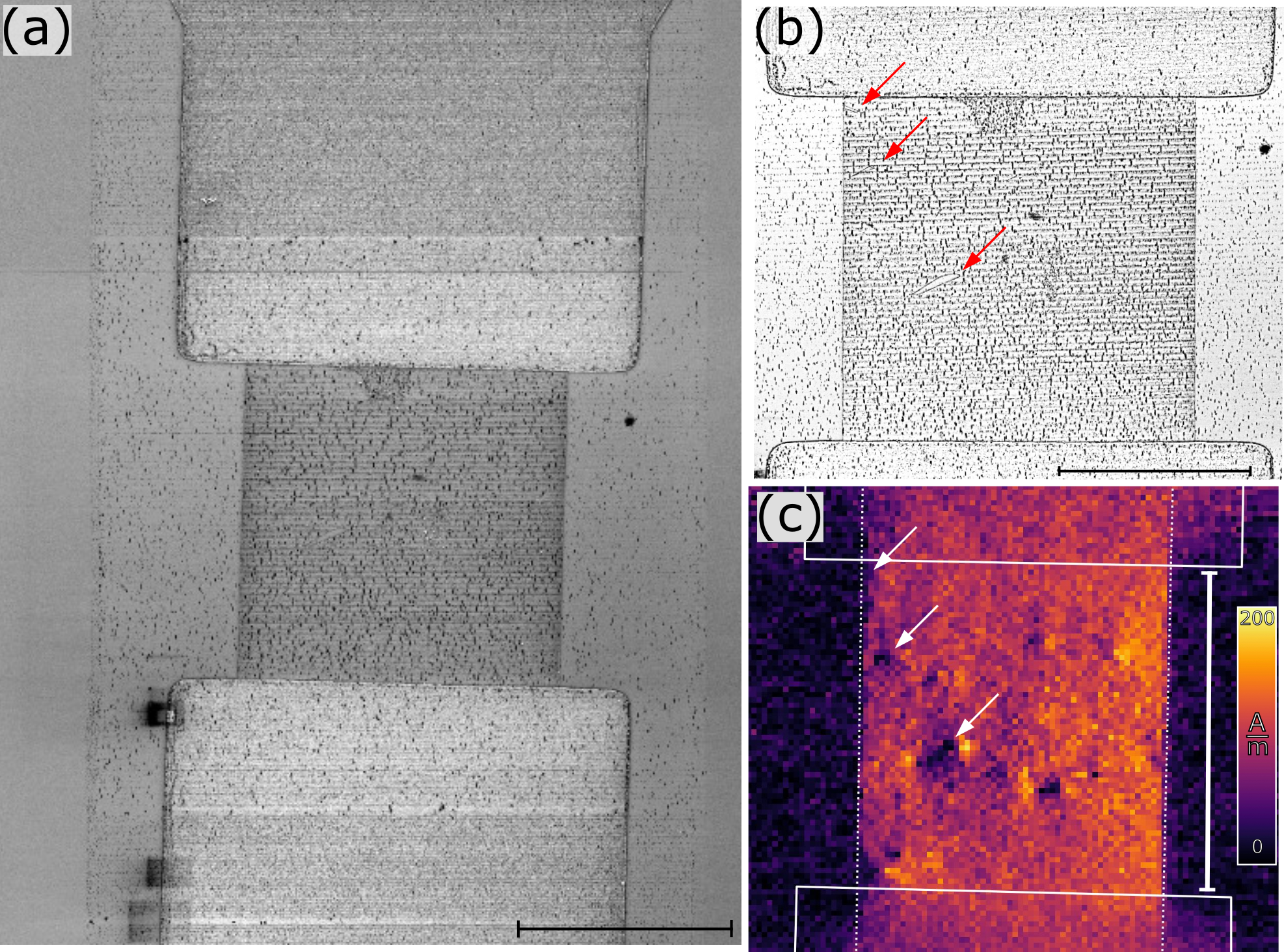}
    \caption{\textbf{Scanning electron microscopy characterization.} Scanning electron microscope (SEM) images of gFET 2, featured in Figure \ref{fig:channel2}: 
    (a) SEM image of the entire device including contact areas. Scale bar is 20 $\upmu$m. (b) Close-up on the device channel with adjusted contrast and brightness settings to increase visibility of graphene defects. Identifiable defects in the channel are highlighted using red arrows. Scale bar is 20 $\upmu$m. (c) Current density map adopted from Figure \ref{fig:channel2}a, cropped and rotated to match the orientation of the SEM image. White arrows indicate the locations of the physical defects in the graphene channel. Scale bar is 30 $\upmu$m.}
    \label{supfig: SEM}
\end{figure}

Following the NV-magnetometry measurements, we characterized gFET 2 in Figure \ref{fig:channel2} using scanning electron microscopy (SEM). The results are shown in Figure \ref{supfig: SEM}: Panel (a) shows an overview of the entire device, capturing a significant portion of the contact regions. Panel (b) provides a closer view of the channel area with adjusted contrast and brightness to increase the visibility of defects in the graphene layer, highlighted by red arrows. In Panel (c), a cropped version of the current density map taken from Figure \ref{fig:channel2}a is shown on which the same defects are highlighted using white arrows. The physical defects in the graphene layer highlighted in the SEM image in Figure \ref{supfig: SEM}b correspond to dark spots on the current density map in Figure \ref{supfig: SEM}c.

In general, Figure \ref{supfig: SEM} shows grooves and residues within the area of the device where the NV-magnetometry measurements were performed. Their absence in other parts of the devices highlights the fact that they are created during the measurement process. The effect of the grooves and residues on the measurement results is further investigated quantitatively by AFM (Figure \ref{supfig: AFM_height}) and a numerical simulation relating the height of the NV-tip to the measured magnetic field (Figure \ref{supfig: height_simulation}); and qualitatively by looking at repeated, non-consecutive measurements of the same device. 
\newpage
\subsection*{Tapping-mode AFM}
\begin{figure}[H]
    \centering
    \includegraphics[width= 0.9\linewidth]{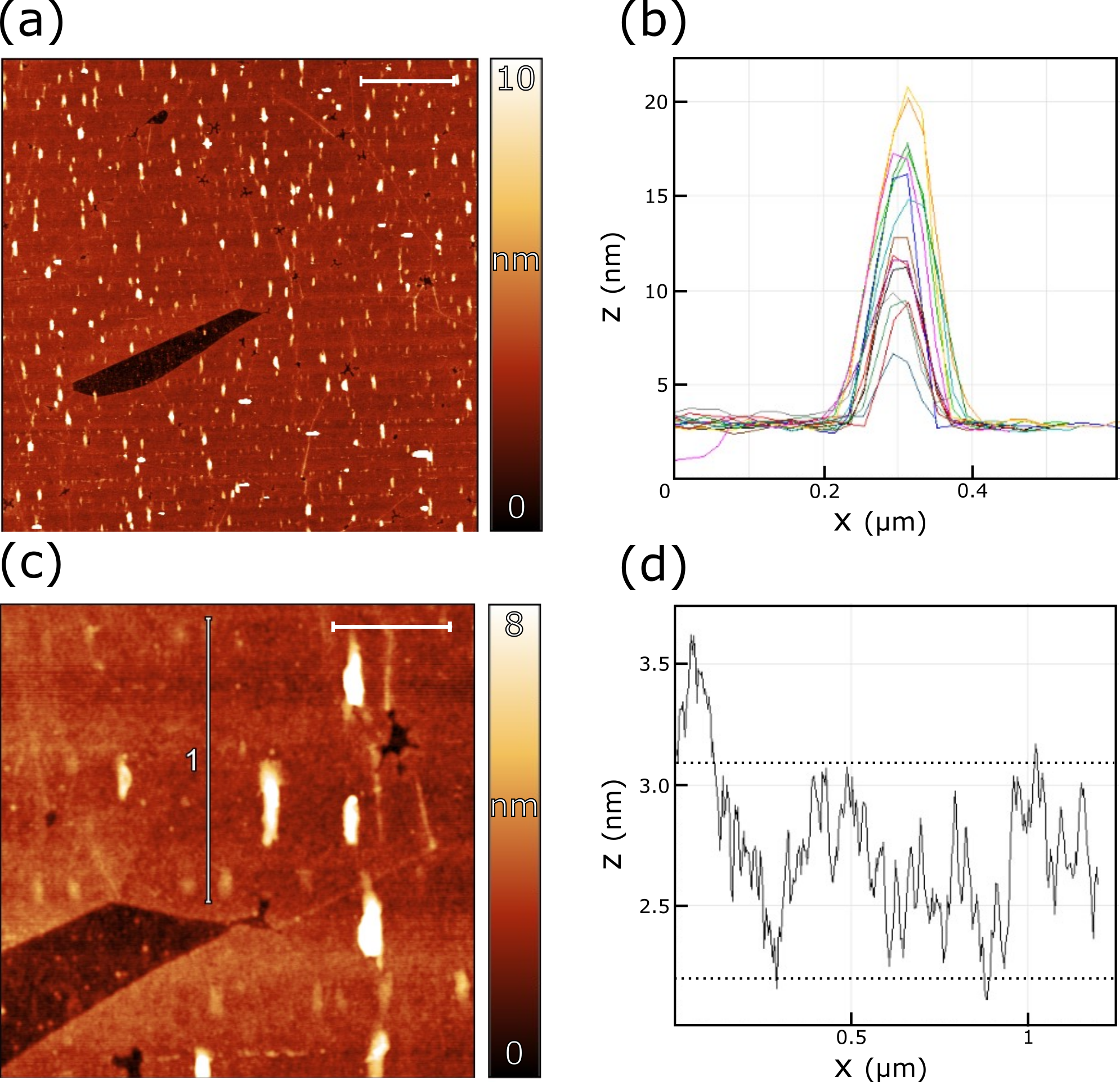}
    \caption{\textbf{Tapping-mode atomic force microscopy characterization.} Atomic force microscopy (AFM) images and corresponding height data of gFET 2, featured in Figure \ref{fig:channel2}: (a) $10~\upmu \mathrm{m} \times 10~\upmu \mathrm{m}$ image above the graphene channel highlighting the presence of high ``residues" on top of the ALD layer. (b) Height analysis of the visible residues, acquired by measuring the height profile of 15 randomly selected residues on the figure in panel (a). Measured heights are between approx. 3 nm and 18 nm. (c) Zoomed in, $5~\upmu \mathrm{m} \times 5~\upmu \mathrm{m}$ AFM image of the same region revealing the grooves made by the NV tip during the magnetometry measurements as horizontal lines. (d) Height profile along the line traced in panel (c), denoted with the number 1. The height profile contains a periodic signal, between the dotted lines, with a peak-to-peak value of roughly 1 nm.}
    \label{supfig: AFM_height}
\end{figure}

Figure \ref{supfig: AFM_height} presents tapping-mode AFM measurements conducted using a Bruker FastScan system on gFET 2, featured in Figure \ref{fig:channel2}. These AFM measurements were performed after all NV-magnetometry and SEM measurements. The purpose of the AFM data is to quantify the groove depths within the ALD layer and the heights of the residues observed in the SEM images of Figure \ref{supfig: SEM}. A $10~\upmu \mathrm{m} \times 10~\upmu \mathrm{m}$ AFM image of the device channel area is shown in Figure \ref{supfig: AFM_height}a, clearly revealing the presence of residues. In Figure \ref{supfig: AFM_height}b, we trace the height of 15 randomly selected residues and find that they measure between 3 nm and 18 nm in height. From Figure \ref{supfig: AFM_height}c, a zoomed-in $5~\upmu \mathrm{m} \times 5~\upmu \mathrm{m}$ AFM image, we can extract the depth of the scratches by extracting the height along a vertical line-trace, denoted by the number 1. The height along this trace is shown in Figure \ref{supfig: AFM_height}d, revealing a periodic signal (between the dotted lines) with a peak-to-peak value of roughly 1 nm. 
\newpage
\subsection*{Height versus magnetic field simulation}
\begin{figure}[H]
    \centering
    \includegraphics[width=\linewidth]{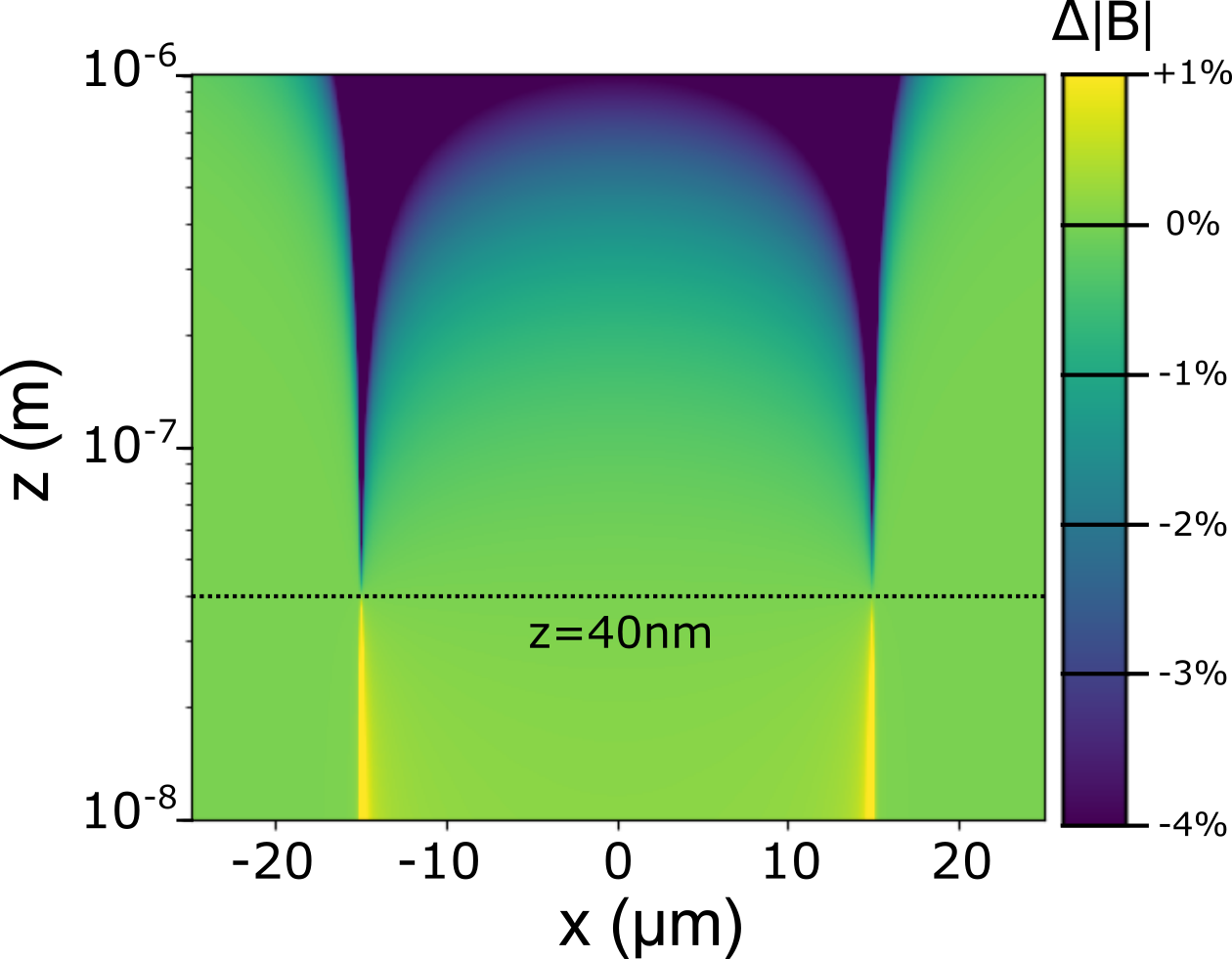}
    \caption{\textbf{Simulated correlation between NV-tip height and magnetic field.} Simulation of the change in magnetic field above an infinite wire carrying current. The wire is 30 $\upmu$m wide, located between $-$15 $\upmu$m and +15 $\upmu$m on the x-axis. The length of the wire extends infinitely in the y-direction, perpendicular to the image plane. The color represents the change in absolute magnetic field with respect to 40 nm above the center of the wire ($x = 0 \ $$\upmu$m, $z = 40\ $nm). The change in magnetic field above the center of the wire reaches $-$1 $\%$ when the height increases from 40 nm to 275 nm. }
    \label{supfig: height_simulation}
\end{figure}

The grooves and residues revealed by the SEM and AFM images in Figures \ref{supfig: SEM} and \ref{supfig: AFM_height} may have a potential impact on the NV-magnetometry results by locally altering the height of the NV-tip relative to the graphene layer. To evaluate the significance of this effect, we simulate the variation in magnetic field above a current-carrying wire as a function of height. The width of the wire is equal to the width of the graphene channels, i.e. 30$\upmu$m. The NV center is located approximately 10 nm inside the tip, and the thickness of the ALD layer is 20 nm. An extra 10 nm of space is added to take into account variations in the tip height due to the feedback response of the instrument. As a result, we simulate the magnetic field variation with respect to the center of a metallic wire at a height of 40 nm. The simulation result is shown in Figure \ref{supfig: height_simulation}. For a height increase from 40 nm to 275 nm, we record a 1 $\%$ decrease in magnetic field. Therefore, we expect topography differences of up to 18 nm to have negligible effects on the NV-measurements, according to this simulation less than 0.08 $\%$.  
\newpage
\subsection*{Repeatability}
\begin{figure}[H]
    \centering
    \includegraphics[width=\linewidth]{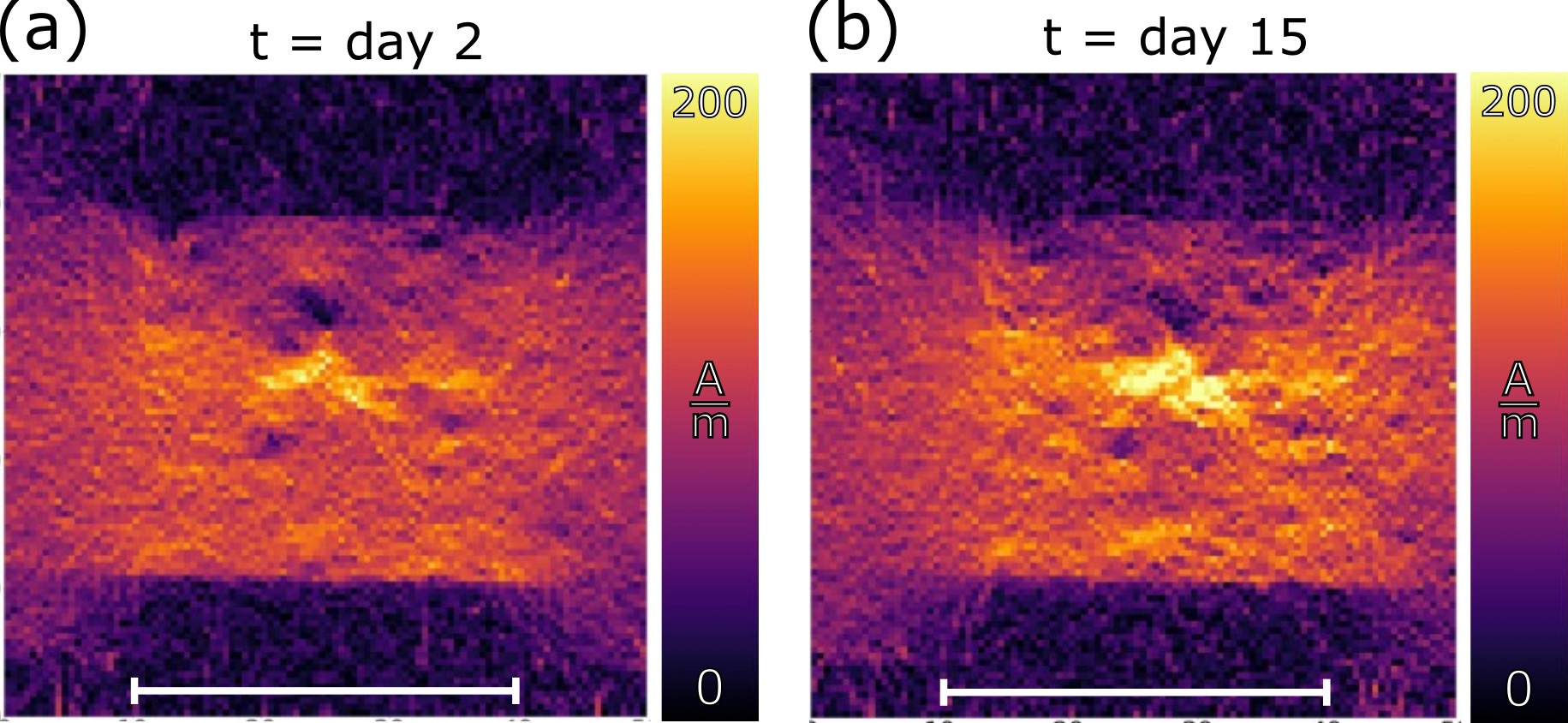}
    \caption{\textbf{Measurement repeatability.} Current density maps of the same region of gFET 2, featured in Figure \ref{fig:channel2}, measured at different times demonstrating measurement repeatability:  
    (a) density map of device channel taken at t = 2 days. (b) density map of the same channel taken at t = 15 days. Identical measurement settings were used for both measurements.$V_g$ = 0 V. Scale bars are 30 $\upmu$m.}
    \label{supfig: repeatability}
\end{figure}

If the grooves and residues on the surface of the gFETs are created during the measurement procedure, they are unlikely to be identical for each measurement. Therefore, analyzing the repeatability of the measurements is a qualitative way to asses their impact on the measurement results. Apart from the the scratches and residues, graphene degradation, shift in Dirac point, environmental differences and setup instability can all impact the measurement repeatability. In Figure \ref{supfig: repeatability}, identical measurements are performed 13 days apart on gFET 2, featured in Figure \ref{fig:channel2}. 
\newpage
\subsection*{Stability of the graphene transfer curve}
\begin{figure}[H]
    \centering
    \includegraphics[width=\linewidth]{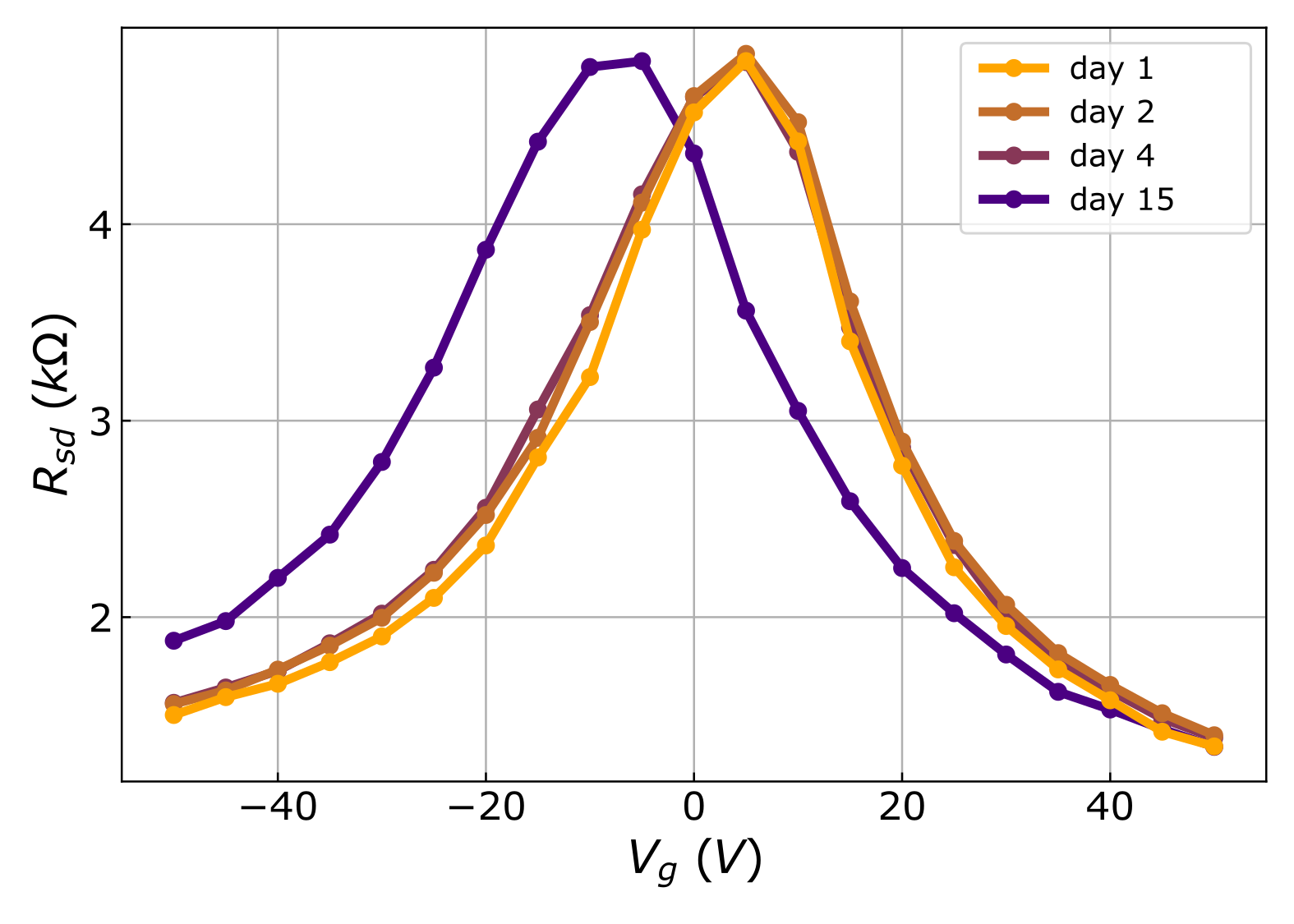}
    \caption{\textbf{Electrical stability of graphene.} Evolution of the transfer characteristic of gFET 2, featured in Figure \ref{fig:channel2}, during the NV-magnetometry measurements. Data is acquired with an applied bias current of +1.5 mA while sweeping the gate from $-$50 V to +50 V. }
    \label{supfig: EL_stability}
\end{figure}

Figure \ref{supfig: EL_stability} shows measured transfer curves of gFET 2 (featured in Figure \ref{fig:channel2}) during the NV-magnetometry measurements. The transfer curves are obtained by applying a bias of +1.5 mA while sweeping the gate (285 nm $\mathrm{SiO_2}$ dielectric) from $-$50 V to 50 V. The NV-magnetometry measurements of gFET 2 spanned a period of 15 days, during which the setup was regularly calibrated, and all density maps featured in this study were acquired. 

In Figure \ref{fig:channel2} we observe that the transfer curve remained stable for the first four days but the Dirac point shifted to the left by roughly 10 V between day 4 and 15. This shift is unlikely to be caused by pinholes in the ALD layer, as the samples were exposed to ambient conditions for several days prior to the measurements without significant changes. Instead, a plausible explanation is the gradual degradation of graphene due to heating from high bias currents. This small voltage shift on such a thick dielectric guarantees that the measured device retains a well-defined electron and hole regime, with a consistent on-off ratio during all NV-magnetometry measurements.  

\newpage
\subsection*{Qualitative comparison with a resistor network model}
Here we simulate the current flow from the gold contacts into the graphene layers using an approach adopted from \cite{mishra2020effective}. The model consists of a finite set of points arranged within a lattice. Each lattice point is connected to its neighbors through resistances, forming a network. A voltage is applied at one end of the lattice, and Kirchhoff’s current law is used to calculate the voltages at each lattice point. Ohm’s law is then employed to compute the current between any two connected lattice points.

The overall device is structured into two overlapping sublattices: one representing the gold layer and the other representing the graphene layer. The gold layer sublattice is 40 $\upmu$m wide, while the graphene sublattice is 30 $\upmu$m wide. Both sublattices extend 60 $\upmu$m in length, with an overlap region spanning 30 $\upmu$m. The distance between adjacent lattice points in both layers is 0.5 $\upmu$m. In the regions where the gold and graphene layers overlap, resistances connect the lattice points in the two sublattices.

The model employs three distinct resistance values to represent different regions within the system. The resistance in the gold layer, the resistance in the graphene layer, and the resistance connecting the overlapping layers. To make the model resemble the experimental measurements more closely, noise is introduced. This noise is derived from the measured data and follows a normal distribution centered around the measured current values.

When comparing the simulation to the experimental data, the current flow shows a qualitative correspondence between the measured and simulated results.  
\begin{figure}[H]
    \centering
    \includegraphics[width=0.5\linewidth]{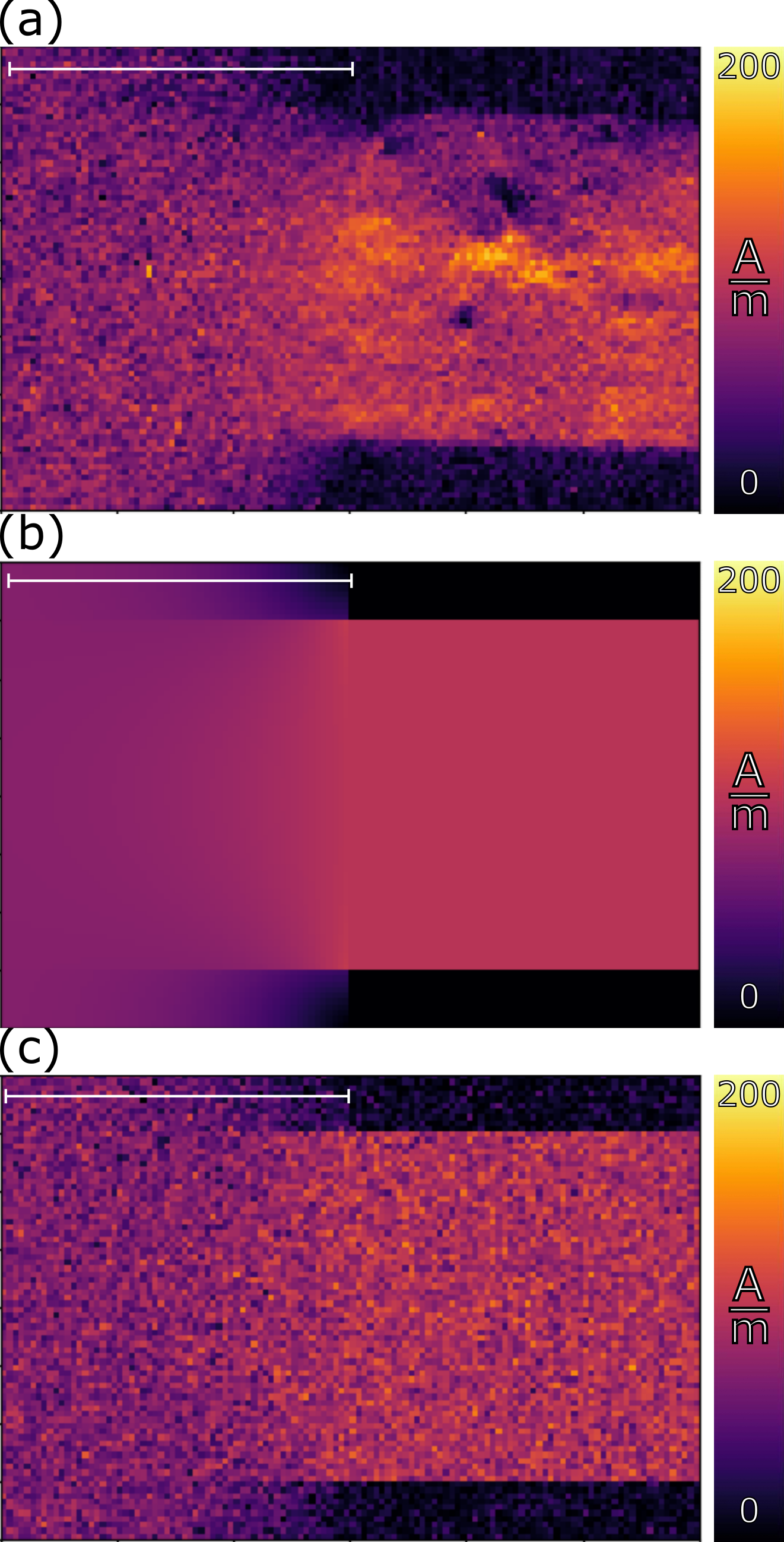}
    \caption{\textbf{Comparison between measured and simulated current density maps.} The current density in gFET 2 is simulated using an approach adopted from \cite{mishra2020effective}, featuring two sublattices (one for the gold and one for the graphene layer) where resistances connect the points between the two sublattices. (a) Measured current density map using the NV-magnetometry setup. (b) Simulated current density map without noise (c) Simulated current density with noise derived from the measured data following a normal distribution. Scale bars are 30 $\upmu$m.}
    \label{supfig: simulation}
\end{figure}
\end{document}